\documentclass[preprintnumbers,amsmath,amssymb,nofootinbib, prd, preprint]{revtex4}
\pdfoutput=1
\usepackage{graphicx,cancel,amssymb}
\usepackage{amsmath,dsfont,textcomp}
\usepackage{epsf}
\usepackage{CJK}

\begin{document}


\title{Two fields inflation with non-minimal coupling}
\author{Jinsu Kim, Yoonbai Kim, Seong Chan Park}
\affiliation{
Department of Physics, BK21 Physics Research Division,
and Institute of Basic Science\\
Sungkyunkwan University, Suwon 440-746, Korea}
\email{kimjinsu@skku.edu, yoonbai@skku.edu, s.park@skku.edu}


\begin{abstract}
Motivated by the recent `Higgs-inflation' scenario based on a single inflaton field, we consider more generic two-field inflation with non-minimal coupling term. The generic analytic expressions are derived  for cosmological observables with the product-separable as well as additive-separable potentials when the non-minimal coupling term is dominated by one of the two fields. 
A hybrid model with the inflaton potential $V=\mu^2 \phi^2 \left[1+\cos(\chi/\sigma)\right]$ is closely examined as a concrete example. Compared to the minimal model, the non-minimal model is shown to provide better fit to the recent cosmological observation by WMAP9 and Planck2013 with the relatively lower field values during the inflationary epoch. Most interestingly, a small value of tensor-to-scalar ratio  requires a large non-minimal coupling in our scenario. The model produces non-observably small non-Gaussianity in most of parameter space while a large non-Gaussianity($\sim {\mathcal O}(10)$) is obtainable only when the inflation takes place  in a limited field space along the top of the potential.
\end{abstract}


\maketitle



\section{Introduction \label{sec:intro}}

The standard slow-roll inflationary paradigm, which can give rise to the primordial perturbations, has been emerged as a solution of cosmological problems, namely,  flatness, homogeneity, and isotropy problems \cite{inflation}.  Observational precision has been dramatically improved recently~\cite{WMAP7, WMAP9, Planck1, Planck2, Planck3} and the refined understanding of the origin of inflation would be hoped even though there exists degeneracy problem \cite{EP}. 

Some of the key measurements toward the origin of the inflation are the scalar spectral index, $n_{\zeta}$, the tensor-to-scalar ratio, $r$, and the primordial non-Gaussianity (NG), $f_{\rm{NL}}$.  The recent Planck data combined with the final nine-year WMAP data \cite{Planck1, Planck2} suggest that these observables are within a window:
\begin{eqnarray}
n_{\zeta}&=&0.9603\pm 0.0073 \quad (68\%\text{ C.L.}), \cr
r&<&0.11\quad (95\%\text{ C.L.}),\cr
f^{\text{local}}_{\rm{NL}}&=& 2.7\pm 5.8\quad (68\%\text{ C.L.}).
\end{eqnarray}
Combination of these measurements provide a powerful guidance to distinguish a specific inflationary model from the others. 
A large NG is not expected in single field models~\cite{single}, and thus observations of the large NG would strongly disfavor single field models.

Recently, `Higgs-inflation' scenario where the inflaton, $\phi$,  is identified with the electroweak Higgs boson in Jordan frame with  non-minimal (NM) coupling, $\xi \phi^{2} R$, with the Ricci scalar, $R$, \cite{BS} and its generalization to arbitrary monotonic function, $\xi \phi^{2} R\rightarrow K(\phi) R$, have been studied with one-field model~\cite{PY}. The most interesting observation would be the fact that an asymptotically flat potential in Einstein frame is automatically obtained when the ratio between the scalar potential $V(\phi)$ and the square of the NM coupling term $K(\phi)$ approaches a constant,
\begin{eqnarray}
\frac{V}{K^2} \to C, \qquad \phi \gg 1,
\label{eq:GAB}
\end{eqnarray}
where $C$ is a positive constant.\footnote{It has been shown that the inflationary scenario with NM coupling term could be natural in the presence of hyperbolic extra dimensions \cite{KP}.} The rather extensive study on NG in single-field inflation model with NM coupling term was recently performed~\cite{Qiu}, in particular, for the $\lambda \phi^4$ potential with $\xi \phi^2 R$ term \cite{Komatsu:1997, Komatsu:1999}.

In this paper we consider two inflaton fields aiming toward even further generic multi-field cases with NM couplings.
Indeed multiple number of scalar fields and their NM couplings are naturally introduced  in generic supergravity theories taking  generic K\"ahler potential into account \cite{Ferrara:2010yw}. 
To be specific, we focus on the two-field inflaton potentials of the form (i) $U_1(\phi)U_2(\chi)$ (product-separable) and (ii) $U_1(\phi)+U_2(\chi)$ (sum-separable) in the presence of the NM coupling term dominated by one of the two fields, namely  $K(\phi,\chi) \approx K(\phi)$ and develop detailed analytic expressions for cosmological observations using $\delta$N formalism.  We reserve more generic cases for the future.

The rest of this paper is organized as follows: In section \ref{sec:2fields}, we develop a set of tools in $\delta $N-formalism to analyze the two-field inflation with NM coupling term. The analytic expressions for physical quantities including power spectrum, spectral index, tensor-to-scalar ratio, and non-linearity parameter are derived for a generic setup. In the subsequent section \ref{sec:Model}, as a concrete example, we apply our results to a specific model of separable potential, $V=\mu^2 \phi^2 [1+\cos (\chi/\sigma)]$ with the NM term $K=\alpha \phi$. For the future experiment, we show the parameter space which is consistent with the current observation. Finally, the conclusion is given in section \ref{sec:conclusion}. In appendix, we provide useful analytic expressions for cosmological observables in various two-field cases with and without NM couplings, which are extensively used in the text.


\section{General analysis of inflation with Two Fields and Non-minimal Couplings \label{sec:2fields}}

Having two scalar fields, $(\phi_{\rm{J}}, \chi_{\rm{J}})$, we write the generic (super)gravity action in Jordan frame as 
\begin{eqnarray}
\frac{{\cal L}_{\rm{J}}}{\sqrt{-g^{\rm{J}}}}=-\frac{1+K(\phi_{\rm{J}}, \chi_{\rm{J}})}{2}R+\frac{1}{2}(\partial \phi_{\rm{J}})^{2}+\frac{1}{2}(\partial \chi_{\rm{J}})^{2}-V(\phi_{\rm{J}},\chi_{\rm{J}}),
\end{eqnarray}
where the NM coupling term, in general, is expanded as
\begin{eqnarray}
K= K_0 + K_\phi \phi_{\rm{J}} + K_\chi \chi_{\rm{J}}  + K_{\phi\phi} \phi_{\rm{J}}^2+K_{\phi\chi} \phi_{\rm{J}}\chi_{\rm{J}} + K_{\chi\chi} \chi_{\rm{J}}^2 +\cdots
\end{eqnarray}
and $V(\phi_{\rm{J}},\chi_{\rm{J}})$ is the inflaton potential in the Jordan frame.\footnote{Note that we are using $M_{\rm{P}}=1/\sqrt{8\pi G}=1$ units.} The constant term $K_0$ can be absorbed in the definition of $M_{\rm{P}}$ so is neglected in below. The specific forms of  $K$ and $V$ depend on their origins.

For the given NM coupling term, we always find the conformal transformation leading to the Einstein frame:
\begin{eqnarray}
g_{\mu\nu}\rightarrow g_{\mu\nu}^{\rm{E}}=\Omega^{2}g_{\mu\nu},
\end{eqnarray}
where the conformal factor, $\Omega^{2}$, is given by
\begin{eqnarray}
\Omega^{2}=1+K(\phi_{\rm{J}},\chi_{\rm{J}}).
\end{eqnarray}
The resultant action in the Einstein frame is, then,
\begin{eqnarray}
\frac{{\cal L}_{\rm{E}}}{\sqrt{-g^{\rm{E}}}}=-\frac{R^{\rm{E}}}{2}+\frac{\Omega^{-2}}{2}\Big[k_{1}(\partial^{\rm{E}}\phi_{\rm{J}})^{2}+k_{2}(\partial^{\rm{E}}\chi_{\rm{J}})^{2}+k_{3}(\partial^{{\rm{E}}\mu}\phi_{\rm{J}})(\partial^{\rm{E}}_{\mu}\chi_{\rm{J}})\Big]
-U(\phi_{\rm{J}},\chi_{\rm{J}}),
\label{eq:LE}
\end{eqnarray}
where
\begin{eqnarray}\label{eq:K123}
k_{1}&\equiv& 1+\frac{3}{2}\Omega^{-2}\left(K_{,\phi_{\rm{J}}}\right)^{2}, \nonumber \\
k_{2}&\equiv& 1+\frac{3}{2}\Omega^{-2}\left(K_{,\chi_{\rm{J}}}\right)^{2}, \\
k_{3}&\equiv& 3\Omega^{-2}K_{,\phi_{\rm{J}}}K_{,\chi_{\rm{J}}}, \nonumber
\label{eq:ks}
\end{eqnarray}
and $U$ is the potential in the Einstein frame, which is related to $V$ by
\begin{eqnarray}
U=\frac{V}{(1+K)^{2}}.
\label{eq:EP}
\end{eqnarray}
Note that the subscript comma `,' denotes the partial derivative, \textit{e.g.}, $K,_{\phi_{\rm{J}}} \equiv \partial K/{\partial \phi_{\rm{J}}}$.

Now a few comments on the general properties of the action are in order:
\begin{itemize}
\item Up to this point, the form of the Lagrangian is symmetric about $\phi_{\rm{J}} \longleftrightarrow \chi_{\rm{J}}$ if $K(\phi, \chi) = K(\chi,\phi)$ and $V(\phi,\chi)=V(\chi,\phi)$. The symmetry can be (explicitly) broken either by the NM coupling term and/or the potential. 
\item Suppose that $K \gg 1$ and $V \gg 1$ in the limit of large fields, $\phi_{\rm{J}}, \chi_{\rm{J}} \gg 1$. A generic condition for large field inflation can be easily noticed since the potential in the Einstein frame, Eq.~\eqref{eq:EP}, looks
\begin{eqnarray}
\lim_{\phi_{\rm{J}},\chi_{\rm{J}}\to\infty} U = \frac{V}{K^{2}}.
\end{eqnarray}
If the asymptotic value is constant, as in Eq.~\eqref{eq:GAB}, the potential involves a large plateau which  may be responsible for  large field inflation. Even when one of the two directions is asymptotically flat, it is still possible to have a slow-roll inflation along that direction. In this case, the condition for the flat potential at the particular field direction, $\phi_{\rm{J}}$ here, becomes
\begin{eqnarray}
\lim_{\phi_{\rm{J}}\rightarrow\infty}\frac{V}{K^{2}}=\text{Const.}
\end{eqnarray}
These results are analogous to those in \cite{PY}.
\end{itemize}


 The kinetic term in Eq.~\eqref{eq:LE}  is not separable in general and thus is not canonically normalizable as far as $k_1$, $k_2$, and $k_3$ in Eq.~\eqref{eq:K123} are mixed functions of $\phi_{\rm{J}}$ and $\chi_{\rm{J}}$. We restrict our interest to the cases of canonically normalizable two fields. To be specific, considering that the NM coupling is only a function of one of the two fields, $K(\phi_{\rm{J}}, \chi_{\rm{J}}) = K(\phi_{\rm{J}})$ or the terms depending on $\chi_{\rm{J}}$ are negligibly smaller than the terms with $\phi_{\rm{J}}$, we have $k_1 = k_1(\phi_{\rm{J}})$, $k_2=1$, and $k_3=0$. More concretely,
 \begin{eqnarray}
 k_1 = 1+\frac{3}{2}\frac{\left(K_{,\phi_{\rm{J}}}\right)^{2}}{1+K}.
 \end{eqnarray}
 Then the canonically normalized field  in the Einstein frame is obtained by solving the equation:
 \begin{eqnarray}
 \frac{d\phi_{\rm E}}{ d\phi_{\rm J}} = \sqrt{\frac{k_1}{1+K}}\,.
 \end{eqnarray}
The action in the Einstein frame, Eq.~\eqref{eq:LE}, is reduced to a well known form \cite{VW,CHB}:
\begin{eqnarray}
\frac{{\cal L}_{\rm{E}}}{\sqrt{-g^{\rm{E}}}}&=&\frac{R}{2}+\frac{k_{1}}{2 \Omega^2}(\partial \phi_{\rm{J}})^2 + \frac{1}{2 \Omega^2}(\partial \chi_{\rm{J}})^2 -\frac{V(\phi_{\rm{J}}, \chi_{\rm{J}})}{\Omega^4} \nonumber \\
&=&\frac{R}{2}+\frac{1}{2}(\partial \phi_{\rm{E}})^2 + \frac{e^{2b}}{2}(\partial \chi_{\rm{E}})^2 -U(\phi_{\rm{E}}, \chi_{\rm{E}}),
\label{eq:lag}
\end{eqnarray}
where $\phi_{\rm{E}}$ is canonically normalized field of $\phi_{\rm{J}}$, $\chi_{\rm{E}}=\chi_{\rm{J}}$, and $b =-\frac{1}{2} \log (1+K)$. One should notice that the ordinary minimally coupled case is recovered  by letting $b=0$ (or $K=0$).

 If the potential is a product-separable in Jordan frame, as $V =U_1(\phi_{\rm{J}}) U_2(\chi_{\rm{J}})$, the potential in Einstein frame is also separable as $U = \tilde{U}_1(\phi_{\rm{E}}) U_2 (\chi_{\rm{E}})$ where $\tilde{U}_1(\phi_{\rm{E}}(\phi_{\rm{J}})) \equiv e^{4b}U_1$. This type of model has  been extensively studied in \cite{VW,CHB} and we will closely follow the existing studies in subsequent sections. In addition, for the brevity of the notation, we will drop subscript `$\rm{E}$' in the Einstein frame below.

In the background Friedmann-Robertson-Walker metric, the equations of motion for the scalar fields are
\begin{eqnarray}\label{eom}
\ddot{\phi}+3H\dot{\phi}+U_{,\phi}=b_{,\phi}e^{2b}\dot{\chi}^{2},\nonumber \\
\ddot{\chi}+\left(3H+2b_{,\phi}\dot{\phi}\right)\dot{\chi}+e^{-2b}U_{,\chi}=0,
\end{eqnarray}
where $H$ is the Hubble parameter. From the Einstein equations we get
\begin{eqnarray}\label{Eeq}
H^{2}&=&\frac{1}{3}\left(\frac{1}{2}\dot{\phi}^{2}+\frac{1}{2}\dot{\chi}^{2}e^{2b}+U\right).
\end{eqnarray}
Under the assumption of slow-roll, Eqs.~\eqref{eom}--\eqref{Eeq} become
\begin{eqnarray}
3H\dot{\phi}+U_{,\phi}=0,\qquad
3H\dot{\chi}+e^{-2b}U_{,\chi}=0,\qquad
H^{2}=\frac{1}{3}U.
\end{eqnarray}

In slow-roll inflationary models, it is convenient to introduce slow-roll parameters
\begin{eqnarray}\label{srp}
\epsilon^{\phi} &\equiv& \frac{1}{2}\left( \frac{U_{,\phi}}{U} \right)^{2}, \quad
\epsilon^{\chi} \equiv \frac{1}{2}\left( \frac{U_{,\chi}}{U} \right)^{2}e^{-2b}, \quad
\epsilon = \epsilon^{\phi} + \epsilon^{\chi}, \nonumber \\
\eta^{\phi} &\equiv& \frac{U_{,\phi \phi}}{U}, \quad
\eta^{\chi} \equiv  \frac{U_{,\chi \chi}}{U}e^{-2b}, \quad
\epsilon^{b}\equiv 8(b_{,\phi})^{2}, \quad
\eta^{b}\equiv 16b_{,\phi\phi}.
\end{eqnarray}
The slow-roll condition requires $\epsilon^{\phi}\ll1$, $\epsilon^{\chi}\ll1$, $|\eta^{\phi}|\ll1$, and $|\eta^{\chi}|\ll1$.
It is also useful to introduce a dimensionless angle, $\theta$, defined by
\begin{eqnarray}
\cos\theta=\frac{\dot{\phi}}{\sqrt{\dot{\phi}^{2}+e^{2b}\dot{\chi}^{2}}},\qquad
\sin\theta=\frac{\dot{\chi}}{\sqrt{\dot{\phi}^{2}+e^{2b}\dot{\chi}^{2}}}e^{b},
\end{eqnarray}
and, in the slow-roll approximation,
\begin{eqnarray}
\cos^{2}\theta=\frac{\epsilon^{\phi}}{\epsilon},\qquad
\sin^{2}\theta=\frac{\epsilon^{\chi}}{\epsilon}.
\end{eqnarray}

In terms of the dimensionless angles,  the cosmological observables are expressed as\footnote{Some details of the derivation are given in Appendix.}
%
\begin{eqnarray}\label{eq:COinT}
\mathcal{P}_{\zeta}&=&\left(\frac{H_{*}}{2\pi}\right)^{2}\frac{e^{2X}}{2\epsilon_{*}}\frac{\cos^{4}\theta_{\rm{e}}}{\sin^{2}\theta_{*}}\left(\mathcal{A}^{2}\tan^{2}\theta_{*}+\tan^{4}\theta_{\rm{e}}\right), \nonumber \\
n_{\zeta}
&=&1-2\epsilon_{*}-4\epsilon_{*}\frac{\sin^{2}\theta_{*}}{\cos^{4}\theta_{\rm{e}}}\frac{e^{-2X}}{\mathcal{A}^{2}\tan^{2}\theta_{*}+\tan^{4}\theta_{\rm{e}}}
-\frac{\cos^{2}\theta_{*}}{12}\frac{(\mathcal{A}\tan^{2}\theta_{*}-\tan^{2}\theta_{\rm{e}})^{2}}{\mathcal{A}^{2}\tan^{2}\theta_{*}+\tan^{4}\theta_{\rm{e}}}\left(\eta_{*}^{b}+2\epsilon_{*}^{b}\right)\epsilon_{*}\nonumber\\
&&+2\frac{\mathcal{A}^{2}\tan^{2}\theta_{*}\eta_{*}^{\phi}+\tan^{4}\theta_{\rm{e}}\eta_{*}^{\chi}}{\mathcal{A}^{2}\tan^{2}\theta_{*}+\tan^{4}\theta_{\rm{e}}}+8\epsilon_{*}\frac{\mathcal{A}\tan^{2}\theta_{\rm{e}}\sin^{2}\theta_{*}}{\mathcal{A}^{2}\tan^{2}\theta_{*}+\tan^{4}\theta_{\rm{e}}} \nonumber \\
&&-s^{\phi}s^{b}\sqrt{\epsilon_{*}^{\phi}\epsilon_{*}^{b}}\tan^{2}\theta_{\rm{e}}\frac{2\mathcal{A}\tan^{2}\theta_{*}-\tan^{2}\theta_{\rm{e}}}{\mathcal{A}^{2}\tan^{2}\theta_{*}+\tan^{4}\theta_{\rm{e}}}, \nonumber \\
r&=&16\epsilon_{*}\frac{\sin^{2}\theta_{*}}{\cos^{4}\theta_{\rm{e}}}\frac{e^{-2X}}{\mathcal{A}^{2}\tan^{2}\theta_{*}+\tan^{4}\theta_{\rm{e}}},\\
f_{\rm{NL}}^{(4)}&=&\frac{5}{6}\Bigg[2J_{p}\epsilon_{*}-F_{p}\eta_{*}^{\phi}-G_{p}\eta_{*}^{\chi}
+s^{b}s^{\phi}K_{p}\sqrt{\epsilon_{*}^{b}\epsilon_{*}^{\phi}} \nonumber \\
&&+2H_{p}\bigg(\eta_{\rm{e}}^{\phi}\sin^{2}\theta_{\rm{e}}+\eta_{\rm{e}}^{\chi}\cos^{2}\theta_{\rm{e}}
-\frac{1}{2}s^{b}s^{\phi}\sqrt{\frac{\epsilon_{\rm{e}}^{b}}{\epsilon_{\rm{e}}^{\phi}}}\epsilon_{\rm{e}}\sin^{4}\theta_{\rm{e}}-4\epsilon_{\rm{e}}\cos^{2}\theta_{\rm{e}}\sin^{2}\theta_{\rm{e}}
\bigg)
\Bigg],\nonumber
\end{eqnarray}
where $X\equiv 2b_{\rm{e}}-2b_{*}$, $\mathcal{A}\equiv e^{-X}\left[1+\left(1-e^{X}\right)\tan^{2}\theta_{\rm{e}}\right]$,
\begin{eqnarray}
s^{i=\phi,\chi}=\bigg\{\begin{array}{c}
+1, \text{ if  $U_{,i}> 0$}\\
-1, \text{ if  $U_{,i}< 0$} \\
\end{array}, \quad
s^{b}=\bigg\{\begin{array}{c}
+1, \text{ if  $b^{\prime}> 0$}\\
-1, \text{ if  $b^{\prime}< 0$} \\
\end{array},
\end{eqnarray}
and
\begin{eqnarray}\label{eq:obs}
J_{p} &\equiv& \frac{\sin^{2}\theta_{*}}{\cos^{2}\theta_{\rm{e}}}e^{-X}\frac{\mathcal{A}^{3}\tan^{2}\theta_{*}+\tan^{6}\theta_{\rm{e}}}{(\mathcal{A}^{2}\tan^{2}\theta_{*}+\tan^{4}\theta_{\rm{e}})^{2}},\nonumber \\
F_{p} &\equiv& \frac{e^{-X}}{\cos^{2}\theta_{\rm{e}}}\frac{\mathcal{A}^{3}\tan^{4}\theta_{*}}{(\mathcal{A}^{2}\tan^{2}\theta_{*}+\tan^{4}\theta_{\rm{e}})^{2}},\nonumber \\
G_{p} &\equiv& \frac{e^{-X}}{\cos^{2}\theta_{\rm{e}}}\frac{\tan^{6}\theta_{\rm{e}}}{(\mathcal{A}^{2}\tan^{2}\theta_{*}+\tan^{4}\theta_{\rm{e}})^{2}},\\
K_{p} &\equiv& \frac{e^{-X}}{\cos^{2}\theta_{\rm{e}}}\frac{\mathcal{A}^{2}\tan^{4}\theta_{*}\tan^{2}\theta_{\rm{e}}}{(\mathcal{A}^{2}\tan^{2}\theta_{*}+\tan^{4}\theta_{\rm{e}})^{2}},\nonumber \\
H_{p} &\equiv& \tan^{2}\theta_{\rm{e}}\frac{(\mathcal{A}\tan^{2}\theta_{*}-\tan^{2}\theta_{\rm{e}})^{2}}{(\mathcal{A}^{2}\tan^{2}\theta_{*}+\tan^{4}\theta_{\rm{e}})^{2}}. \nonumber
\end{eqnarray}
Note that the sub(super)script $*(\rm{e})$ indicates the value at the horizon exit (end of inflation) and the prime index denotes the derivative with respect to the corresponding fields as usual.

\begin{figure}[h]\centering
\includegraphics[width=.95\textwidth]{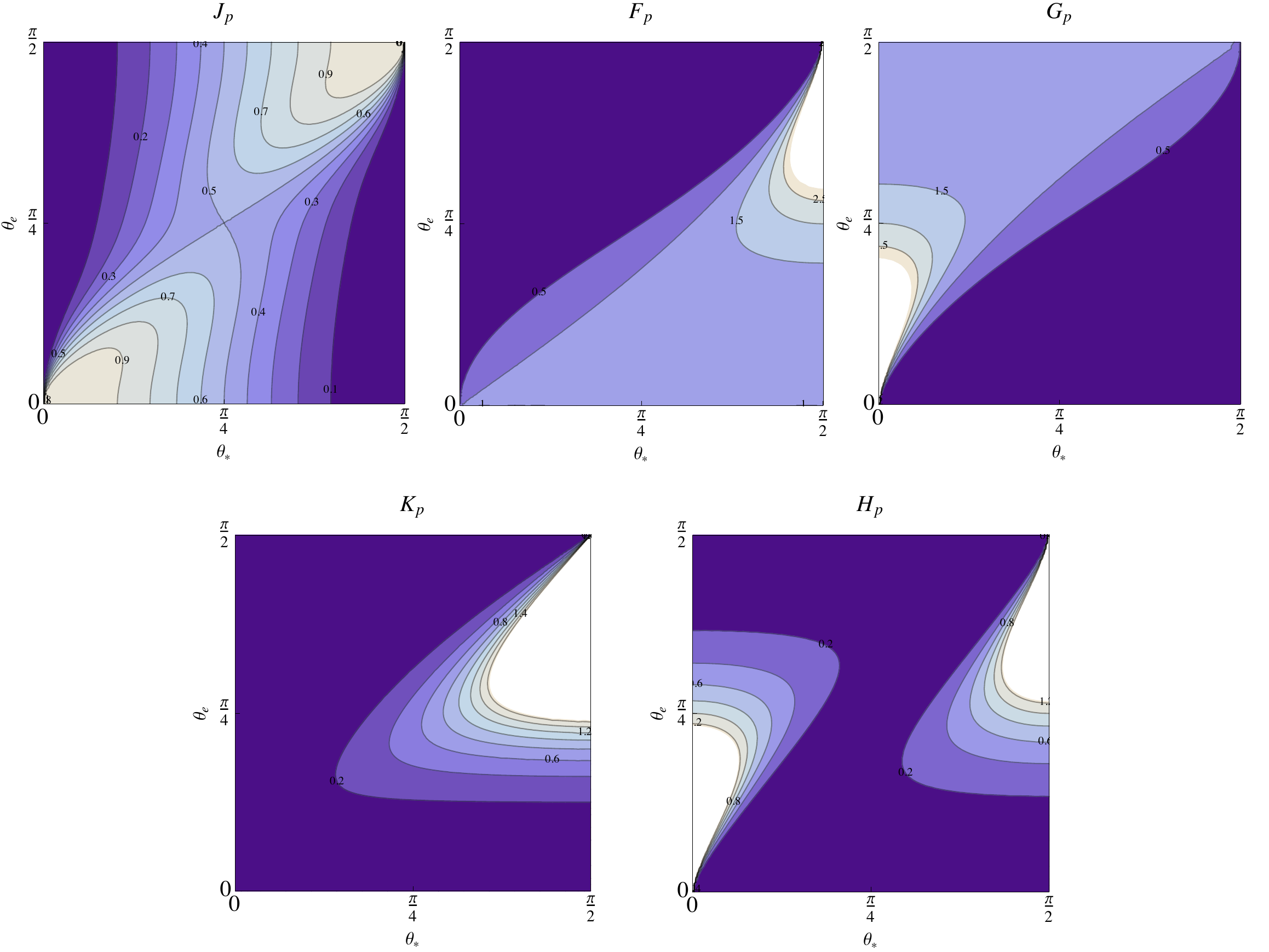}
\caption{\small Analysis of the non-linearity parameter for the non-minimally coupled case with $X=0$. \label{fig:X00}
}
\end{figure}

In order to see general tendency of the non-linearity parameter we plotted $J_p$, $F_p$, $G_p$, $K_p$, and $H_p$ in Fig.~\ref{fig:X00} with fixed $X=0.0$ for numerical evaluation. Similar results are found for $X=\pm 0.1$, too.  The general tendency is summarized as follows:
\begin{itemize}
\item Since the $J_p$ is always less than unity and smaller than other parameters, we will ignore $J_p$ in evaluating cosmological observables.  
\item The parameters, $F_p$, $K_p$, and $H_p$  can be significant when $\epsilon^{\phi}\ll\epsilon^{\chi}$ or $(\theta_{*},\theta_{\rm{e}})\sim (\pi/2,\pi/2)$. On the other hand, when $(\theta_{*},\theta_{\rm{e}})\sim (0,0)$ or $\epsilon^{\phi}\gg\epsilon^{\chi}$, $G_p$ and $H_p$ are significant. 
\item Due to the large $K_p$-dependent term in Eq.~\eqref{eq:obs},  $\epsilon^{\phi}\ll\epsilon^{\chi} (\epsilon^\phi \gg \epsilon^{\chi})$ case produces  large NG when $s^{b}s^{\phi}>0 (<0)$, respectively.  
\end{itemize}

Details of the cosmological observables depend on the explicit realization of the model. In the next section, we will more closely examine a particular case with $K=\alpha \phi$ and $V= \mu^2 \phi^2 [1+\cos (\chi/\sigma)]$, which leads an interesting phenomenology.


\section{An explicit example: $V = \mu^2 \phi^2 [1+\cos (\chi/\sigma)]$}\label{sec:Model}
\begin{figure}[th]\centering
\includegraphics[width=.7\textwidth]{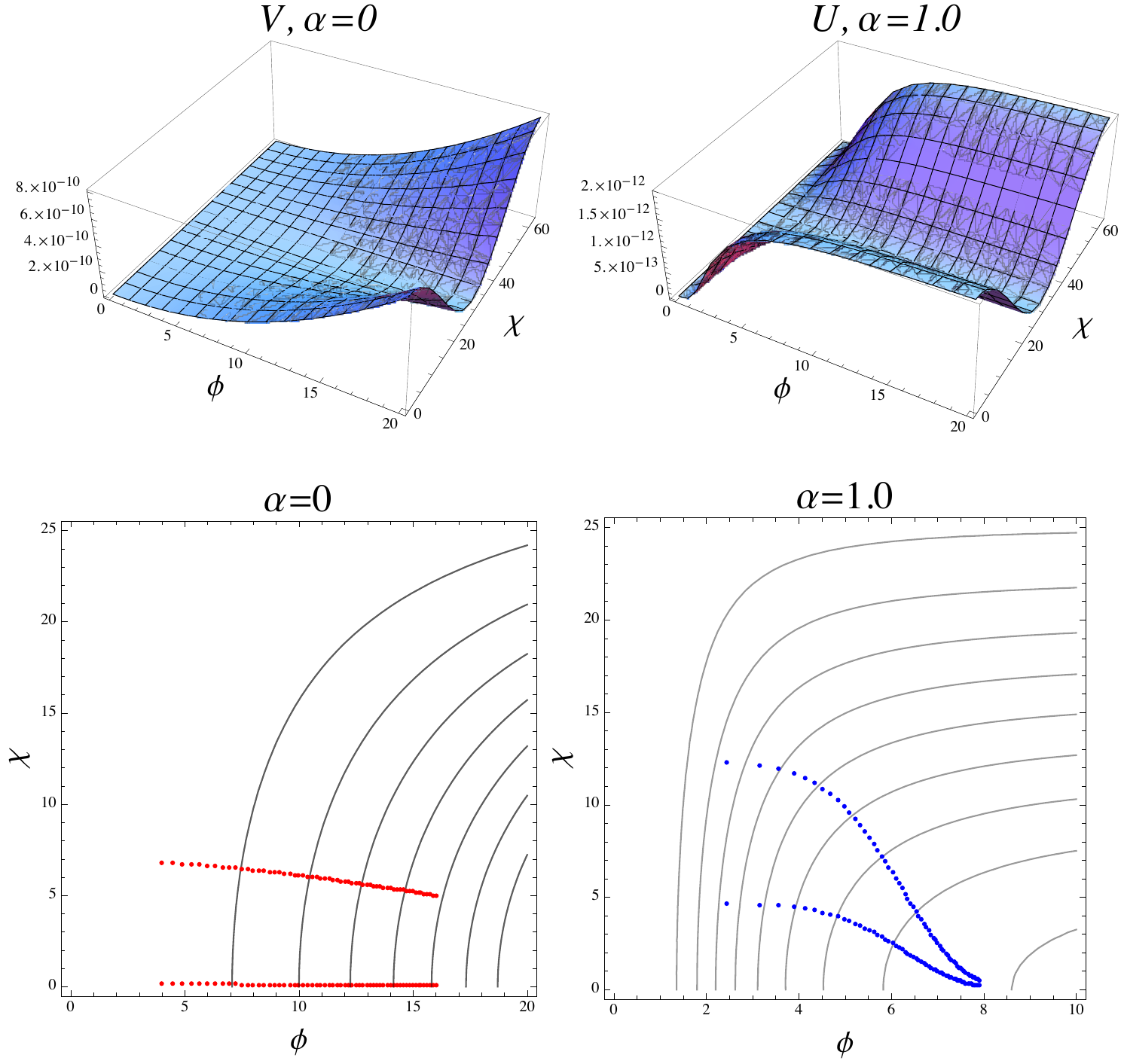}
\caption{Upper two figures are  the inflaton potentials for minimal ($\alpha=0$) and non-minimal ($\alpha=1$) case, respectively. We fix $\mu=1.0\times 10^{-6}$ and $\sigma=10.0$. The two dotted lines in the lower two figures stand for the typical trajectories for minimal ($\alpha=0$) and non-minimal ($\alpha=1.0$) cases with different initial field values. \label{fig:pot}}
\end{figure}

In this section, we consider a phenomenological potential of the form in the Jordan frame,
\begin{equation}
V= \mu^2 \phi^2 \left[1+\cos\left(\frac{\chi}{\sigma}\right) \right].
\label{eq:pot}
\end{equation}
This potential could be regarded as a hybrid of the chaotic inflation with $m^2\phi^2$ term~\cite{chaotic} and the natural inflation with cosine term along $\chi$ direction~\cite{natural}. Interestingly, the potential in the Einstein frame involves a flat plateau along $\phi$ direction (or natural inflation along the orthogonal $\chi$ direction).  When the NM coupling term is of the form of $K = \alpha \phi$, the condition, $V/K \to C$, $\phi \gg 1$  in Eq.~\eqref{eq:GAB} is satisfied.  The potential and the typical paths of inflation are depicted in Fig.~\ref{fig:pot} for minimal ($\alpha=0$, left) and non-minimal ($\alpha\neq 0$, right) case, respectively.

\subsection{Minimally coupled case: $K=0$}
In the minimally coupled case ($\alpha=0$ or equivalently $b=0$ in Eq.~\eqref{eq:lag}), the potential in the Jordan frame and the Einstein frame become identical with $\phi_{\rm J} = \phi$ and $\chi_{\rm J} =\chi$:
\begin{eqnarray}
U(\phi,\chi)&=&\mu^{2}\phi^{2}\left[1+\cos\left(\frac{\chi}{\sigma}\right)\right].
\end{eqnarray}
Here we drop the the subscript, `${\rm E}$'.
The shape of the potential and the typical trajectories of the inflaton are depicted in the left two figures in Fig. \ref{fig:pot}.  The potential may be understood as the mass term along the $\phi$ direction, which varies periodically along the $\chi$ direction.

For each trajectory, the number of e-foldings, $N$, is found in a rather simple form:
\begin{eqnarray}
N=-\frac{1}{4}\left(\phi_{\rm e}^{2}-\phi_{*}^{2}\right)=-2\sigma^{2}\ln\left[\frac{\sin\left(\chi_{*}/2\sigma\right)}{\sin\left(\chi_{\rm e}/2\sigma\right)}\right]
\end{eqnarray}
that determines the evolving field values of $\phi$ and $\chi$ for a given number of e-foldings: 
\begin{eqnarray}
\phi(N)&=&\sqrt{\phi_{*}^{2}-4N},\\
\chi(N)&=&2\sigma\sin^{-1}\left[\sin\left(\frac{\chi_{*}}{2\sigma}\right)\exp\left(\frac{N}{2\sigma^{2}}\right)\right],
\end{eqnarray}
where $\phi_*$ and $\chi_*$ are the initial values of inflaton fields. Here we only consider `large-to-small' evolution along the $\phi$ direction and $\phi_* > 2 \sqrt{N_{60}} \simeq 15.5 $ for reality of $\phi$ during inflation. In general, $\chi$ is a periodic function sitting in the cosine potential but we only restrict ourselves to the case where the inflation takes place within one period of oscillation. 
Taking $N=60$, we find $\phi_{\rm e}$ and $\chi_{\rm e}$ as the values of fields at the end of inflation, 
\begin{eqnarray}
\phi_{\rm e}=\phi(N=60), \quad \chi_{\rm e}=\chi(N=60).
\end{eqnarray}
%

\begin{figure}[h]\centering
\includegraphics[width=.9\textwidth]{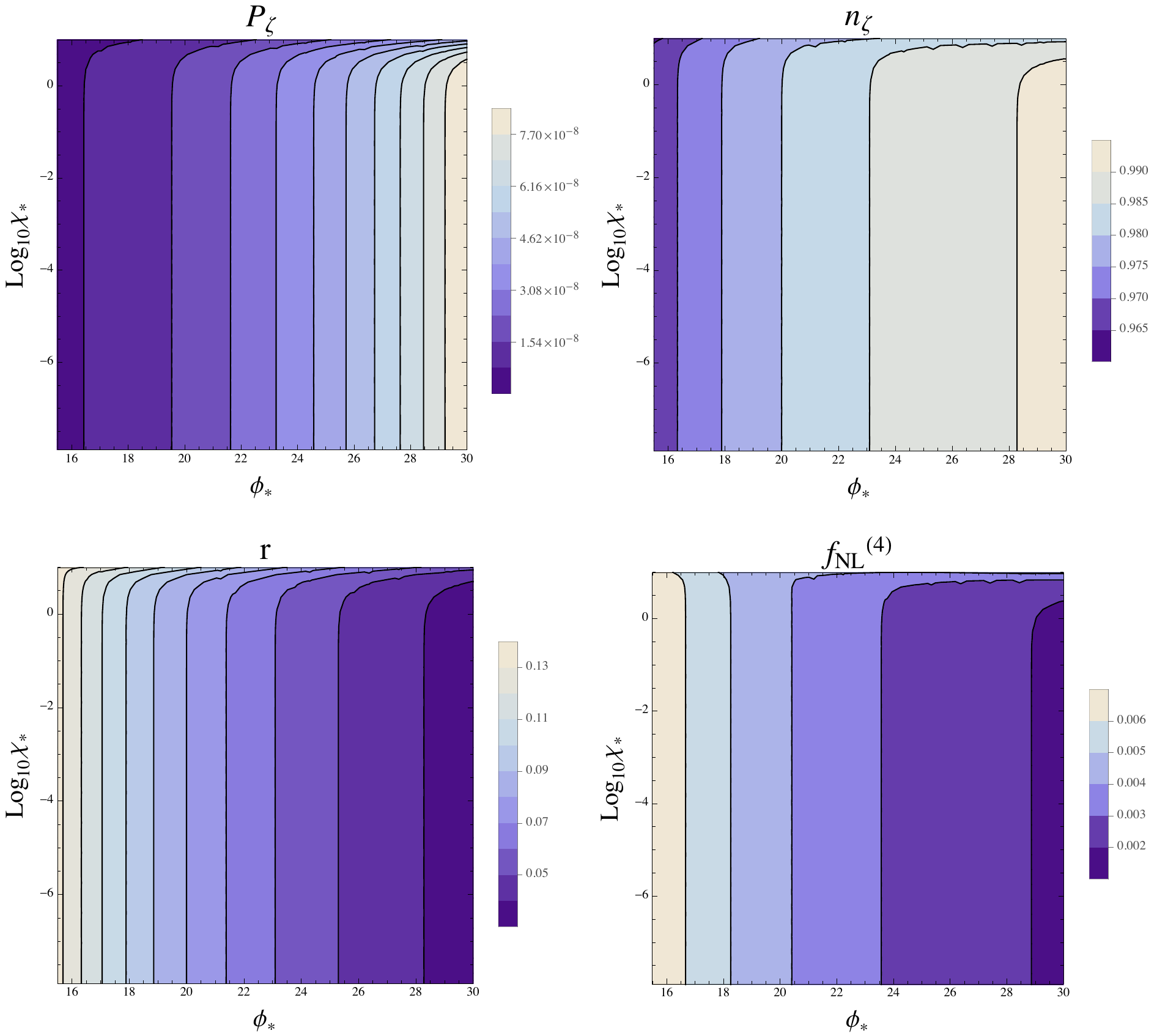}
\caption{Cosmological observables ${\mathcal P}_\zeta$, $n_\zeta$, $r$, and $f_{\rm NL}^{(4)}$ are depicted for $\alpha=0$ case. The large non-Gaussianity above $0.001$ is hardly obtainable in this scenario when the observational constraints from Planck are taken into account.
\label{fig:m}}
\end{figure}

The slow-roll parameters are obtained following the Eq.~\eqref{srp} with $b=0$:
\begin{eqnarray}
\epsilon^{\phi}=\frac{2}{\phi^{2}},\qquad
\epsilon^{\chi}&=&\frac{1}{2\sigma^{2}}\tan^{2}\left(\frac{\chi}{2\sigma}\right), \qquad
\epsilon=\epsilon^{\phi}+\epsilon^{\chi},\nonumber \\
\eta^{\phi}=\frac{2}{\phi^{2}},&\qquad&
\eta^{\chi}=-\frac{1}{\sigma^{2}}\left[\frac{\cos(\chi/\sigma)}{1+\cos(\chi/\sigma)}\right].
\label{eq:min}
\end{eqnarray}
The slow-roll conditions, $\epsilon^{\phi, \chi} \ll 1$, lead to the domain of the field values in 
\begin{eqnarray}
\phi\gg\sqrt{2},\qquad
\chi\ll \sigma\pi.
\end{eqnarray}

Having all the slow-roll parameters in Eq.~\eqref{eq:min}, we can find the analytic expressions of the power spectrum, spectral index, tensor-to-scalar ratio, and the non-linearity parameter~\cite{BCH}. The results are presented in Appendix.
One immediately notices that the power spectrum $\mathcal{P}_{\zeta}\sim \mathcal{O}\left(10^{-9}\right)$ sets the value of $\mu$ parameter as $\mu\sim\mathcal{O}(10^{-6})$ but the other parameter $\sigma$ is still to be determined. Thus we will treat $\sigma$ as a free parameter.

 In Fig.~\ref{fig:m}, we plotted the cosmological observables in $\phi_* \in (15, 30)$ and $\chi_* \in (10^{-8},10^0)$ with $\sigma=10.0$. In the parameter space we typically found $f_{\rm NL}^{(4)}\sim 0.002-0.006$.  Though we did not  present numerical results for a different domain where the inflation takes place with $\chi_* < 10^{-14}$, we sometimes found larger values of $|f_{\rm NL}^{(4)}|>30$. However, this domain turns out to be largely in conflict with the measured cosmological observable: $n_\zeta<0.01$. Also, if the inflation ends with $\epsilon = 1$, the model only produces non-observably small NG, and the values of the tensor-to-scalar ratio, $r$, are turned out to be slightly large, $r>0.1$.
We summarize some representative results in Table~\ref{table:m}.
\begin{table}[ht]
\centering
\begin{tabular}{c c c c c c}
\hline \hline
$\sigma$ & $\phi_{*}$ & $\chi_{*}$ &  $n_{\zeta}$ & $r$ & $f_{\rm{NL}}^{(4)}$ \\ [0.5ex]
\hline \hline
10 & 15.557 & 1.0   & 0.9669 & 0.132 & 0.007\\
10 & 15.560 & 2.0   & 0.9669 & 0.132 & 0.007 \\
10 & 15.560 & 10.0 & 0.9640 & 0.133 & 0.007 \\
10 & 15.570 & 2.0   & 0.9669 & 0.132 & 0.007 \\ [1ex]
\hline
\end{tabular}
\caption{Results for the minimally coupled case with $\alpha=0$ and $\sigma =10$. \label{table:m}
}
\end{table}

\subsection{Non-minimally coupled case: $K=\alpha\phi$ \label{sec:nm}}

Finally, we introduce sizable NM couplings. With the NM coupling term, the inflaton potential in Einstein frame would be different from the one in Jordan frame so that the actual cosmological observables are modified. In Fig.~\ref{fig:pot}  the potential in Einstein frame and the typical evolution of the inflaton fields are depicted for NM case ($\alpha\neq 0$, on right) in contrast to them in the minimal case ($\alpha=0$, on left).

Defining the conformal transformation, $\Omega^{2}=1+\alpha\phi_{\rm J}$, and the $k$'s in  Eq.~\eqref{eq:ks} as
\begin{eqnarray}
k_{1}= 1+\frac{3\alpha^{2}}{2(1+\alpha\phi_{\rm J})},\qquad
k_{2}= 1, \qquad
k_{3}= 0,
\end{eqnarray}
we find the action in Einstein frame following the Eq.~\eqref{eq:LE}:
\begin{eqnarray}
\frac{{\cal L}_{\rm{E}}}{\sqrt{-g^{\rm{E}}}}&=&-\frac{1}{2}R+\frac{1}{2}\frac{1+\alpha\phi_{\rm J}+3\alpha^{2}/2}{(1+\alpha\phi_{\rm J})^{2}}(\partial\phi_{\rm J})^{2}+\frac{1}{2}\frac{1}{1+\alpha\phi_{\rm J}}(\partial\chi_{\rm J})^{2}-\frac{V}{(1+\alpha\phi_{\rm J})^{2}}\nonumber \\
&=&-\frac{1}{2}R+\frac{1}{2}(\partial\phi_{\rm E})^{2}+\frac{1}{2}e^{2b(\phi_{\rm E})}(\partial\chi_{\rm E})^{2}-U(\phi_{\rm E},\chi_{\rm E}),
\end{eqnarray}
where $\chi_{\rm E}=\chi_{\rm J}$ and $\phi_{\rm E}$ is canonically normalized by
\begin{eqnarray}
\frac{d\phi_{\rm E}}{d\phi_{\rm J}}=\frac{\sqrt{1+\alpha\phi_{\rm J}+3\alpha^{2}/2}}{1+\alpha\phi_{\rm J}}.
\end{eqnarray}
The coefficient of the NM kinetic term is
\begin{eqnarray}
e^{2b(\phi_{\rm E})}=\frac{1}{1+\alpha\phi_{\rm J}(\phi_{\rm E})},
\end{eqnarray}
and the inflaton potential in the Einstein frame is 
\begin{eqnarray}
U(\phi_{\rm E}(\phi_{\rm J}),\chi)=\frac{\mu^{2}\phi_{\rm J}^{2}}{(1+\alpha\phi_{\rm J})^{2}}\left[1+\cos\left(\frac{\chi_{\rm E}}{\sigma}\right)\right].
\label{eq:ue}
\end{eqnarray}
Hereafter, we will drop the subscript `${\rm E}$' for brevity of the notation.

We are mostly interested in the large field limit, $\phi_{\rm J}\gg 1/\alpha$, because the potential has a large flat plateau along the $\phi$ direction as we discussed earlier. In this limit, the explicit form of the canonical field is easily obtained as
\begin{eqnarray}
\frac{d\phi}{d\phi_{\rm J}}\approx\frac{1}{\sqrt{\alpha\phi_{\rm J}}} \Longrightarrow
\phi \approx 2\sqrt{\phi_{\rm J}/\alpha} \Longleftrightarrow \phi_{\rm J} \approx \frac{1}{4}\alpha\phi^{2}.
\end{eqnarray}
In the limit, the NM kinetic term looks rather simple,
\begin{eqnarray}
\phi_{\rm J} \gg \frac{1}{\alpha} &\Longleftrightarrow& \alpha^{2}\phi^{2}\gg 4,\qquad e^{2b(\phi)}\approx \frac{1}{\alpha \phi_{\rm{J}}}\approx \frac{4}{\alpha^{2}\phi^{2}},
\end{eqnarray}
and then the potential in Eq.~\eqref{eq:ue} becomes
\begin{eqnarray}
U(\phi,\chi)\approx\frac{\mu^{2}\alpha^{2}\phi^{4}}{(4+\alpha^{2}\phi^{2})^{2}}\left[1+\cos\left(\frac{\chi}{\sigma}\right)\right]\equiv U_{1}(\phi)U_{2}(\chi)
\end{eqnarray}
which leads to
\begin{eqnarray}
U_{1}(\phi)=\frac{\mu^{2}\alpha^{2}\phi^{4}}{(4+\alpha^{2}\phi^{2})^{2}}, \qquad
U_{2}(\chi)=1+\cos\left(\frac{\chi}{\sigma}\right).
\end{eqnarray}
It is easily noticed that the potential becomes flat as $U_1 \to \mu^2 / \alpha^2$ in the large field limit. As $\mu \ll 1$ and we take $\alpha\sim {\cal O}(1)$, the potential is clearly in the sub-Planckian domain.

The slow-roll parameters in Eq.~\eqref{srp} are well approximated:
\begin{eqnarray}
\epsilon^{\phi}&=&\frac{128}{\alpha^{4}\phi^{6}},\quad
\epsilon^{\chi}= \frac{\alpha^{2}\phi^{2}}{8\sigma^{2}}\tan^{2}\left(\frac{\chi}{2\sigma}\right),\nonumber \\
\eta^{\phi} &=& -\frac{48}{\alpha^{2}\phi^{4}}, \quad
\eta^{\chi} = -\frac{\alpha^{2}\phi^{2}}{8\sigma^{2}}\cos\left(\frac{\chi}{\sigma}\right)\sec^{2}\left(\frac{\chi}{2\sigma}\right).
\end{eqnarray}
From the slow-roll conditions for both of the fields, we found the allowed range for $\phi$ and $\chi$ during the inflation as
\begin{eqnarray}
\phi \gg \frac{2^{\tfrac{7}{6}}}{\alpha^{\tfrac{2}{3}}}, \quad
\chi \ll 2\sigma\tan^{-1}\left(\sqrt{\frac{8\sigma^{2}}{\alpha^{2}\phi^{2}}}\right).
\end{eqnarray}

\begin{figure}[h!]\centering
\includegraphics[width=.9\textwidth]{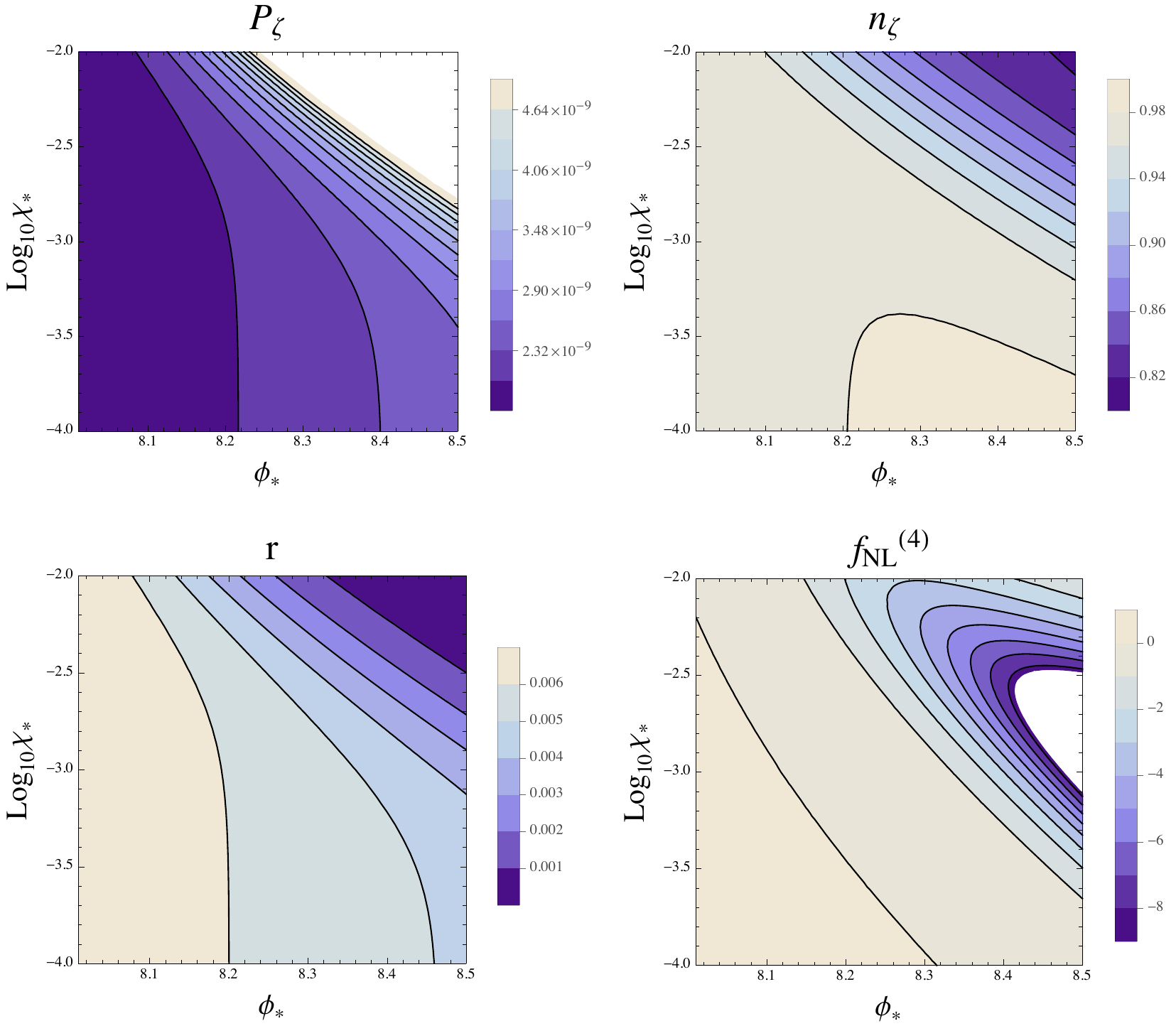}
\caption{Cosmological observables ${\mathcal P}_\zeta$, $n_\zeta$, $r$, and $f_{\rm{NL}}^{(4)}$ are depicted for $\alpha=1.0$ case. The large non-Gaussianity above $|f_{NL}^{(4)}|\geq {\cal O}(10)$ is obtained.
\label{fig:nm}}
\end{figure}

It is convenient to express the evolution of the inflaton fields in terms of the number of e-foldings which is given by
\begin{eqnarray}
N=-\int_{*}^{\rm e}\frac{U_{1}}{U_{1}^{\prime}}d\phi=-\int_{*}^{\rm e}e^{-2b}\frac{U_{2}}{U_{2}^{\prime}}d\chi,
\end{eqnarray}
or more explicitly by 
\begin{eqnarray}
N\approx -\frac{\alpha^{2}}{64}\left(\phi_{\rm e}^{4}-\phi_{*}^{4}\right),
\end{eqnarray}
from which we find
\begin{eqnarray}
\phi(N)\approx \phi_{*}\left(1-\frac{64N}{\alpha^{2}\phi_{*}^{4}}\right)^{1/4}.
\end{eqnarray}
The other field, $\chi(N)$,  is determined by the equation of motion, $3H\dot{\chi}\approx -e^{-2b}U_{,\chi}$,\begin{eqnarray}
\int_{*}^{\chi}\frac{1+\cos(\chi/\sigma)}{\sin(\chi/\sigma)} d\chi \approx \frac{\alpha^{2}}{4\sigma}\int_{0}^{N}\phi^{2}(N)dN 
\approx \frac{1}{4\sigma}\alpha^{2}\phi_{*}^{2}\int_{0}^{N}\left(1-\frac{64N}{\alpha^{2}\phi_{*}^{4}}\right)^{1/2}dN,
\end{eqnarray}
from which we find
\begin{eqnarray}
\chi(N)&\approx&2\sigma\sin^{-1}\left[ \sin\left(\frac{\chi_{*}}{2\sigma}\right)\exp\left\{\frac{\alpha^{4}\phi_{*}^{6}}{96\cdot 8 \sigma^2}\left[1+\Big|1-\frac{64N}{\alpha^{2}\phi_{*}^{4}}\Big|^{3/2}\right]\right\} \right].
\end{eqnarray}
Collecting all the above formulas, we can find cosmological observables. The results are presented in Appendix in a generic form. The numerical results for $P_\zeta$, $n_\zeta$, $r$, and $f_{\rm NL}^{(4)}$ (local) are depicted in Fig.~\ref{fig:nm}. We chose $\mu\sim\mathcal{O}\left(10^{-5}\right)$ to fit  the observed data of the power spectrum of curvature perturbation, $\mathcal{P}_{\zeta}\approx2.5\times10^{-9}$. It is interesting to notice that $\mu\sim 10^{-5}\sim M_{\rm GUT}$, that is the scale of grand unified theory.

One of the most distinctive features of this case with $\alpha=1.0$ is a small value of tensor-to-scalar ratio, $r$. This suppression of $r$ may be understood as an effect of the flat potential in the Einstein frame which is originated from the NM coupling term. In addition, a large NG($\sim\mathcal{O}(10)$) is obtainable when the inflation takes place in a specific field space, $(\phi, \chi)\sim(8.5, 10^{-4})$, and the inflation ends before reaching $\epsilon=1$. On the other hand, if the inflation ends with $\epsilon=1$, the model only produces non-observably small NG.

We show the representative results in Table~\ref{table:nm}.
\begin{table}[ht]
\centering
\begin{tabular}{c c c c c c}
\hline \hline
$\sigma$ & $\phi_{*}$ & $\chi_{*}$ &  $n_{\zeta}$ & $r$ & $f_{\rm{NL}}^{(4)}$ \\ [0.5ex]
\hline \hline
10 & 7.876 & 0.70 & 0.9656 & 0.009 & 0.005 \\
10 & 7.877 & 0.70 & 0.9590 & 0.009 & 0.015 \\
10 & 7.878 & 0.65 & 0.9623 & 0.009 & 0.013 \\
10 & 8.833 & $10^{-4}$ & 0.9647  & 0.003 & -27.67 \\[1ex]
\hline
\end{tabular}
\caption{Results for the non-minimally coupled case with $\alpha=1.0$. \label{table:nm}
}
\end{table}

\section{Conclusion \label{sec:conclusion}}

In a single field inducing inflation, a large non-minimal coupling term of the inflaton field and Ricci scalar gives rise to a flat potential in Einstein frame. This potential is a desirable one in the slow-roll inflationary paradigm \cite{PY} and has been applied to `Higgs-inflation' model \cite{BS}. In this paper, we enlarge the scope of the previous studies by  considering two fields, one of which includes a significant NM coupling. We provide general analytic expressions for physical observables including power spectrum, spectral index, tensor-to-scalar ratio, and non-linearity parameter with product-separable and additive-separable inflaton potentials. These results would be extremely useful for explicit comparison with the existing and forthcoming observational data and a specific choice of inflationary potential. 

As a concrete example, we examine  a hybrid potential, $V=\mu^2 \phi^2 (1+\cos \chi/\sigma)$ assuming that $\phi$ has a large NM coupling. This model is potentially realistic thus deserves further study especially when the NM coupling is involved as is summarized in Table \ref{table:nm}. A small value of tensor-to-scalar ratio requires a large NM coupling (see Table \ref{table:m} for comparison). The model produces a non-observably small NG when it is required that the inflation ends at $\epsilon\sim 1$ in most of parameter space. We also notice that a large NG($\sim {\mathcal O}(10)$) is obtainable when the inflation takes place  in a limited field space, $(\phi, \chi)\sim (8.5, 10^{-4})$, near the top of the inflaton potential.

\section*{Acknowledgments}

This work is supported by the project of Global Ph.D. Fellowship which the National Research Foundation of Korea conducts from 2011 with grant number 2011-0008792 (J. K.) and by Basic Science Research Program through the National Research Foundation of Korea funded
by the Ministry of Education, Science and Technology with grant numbers 2011-0011660 (Y. K.), and also 2011-0010294 and 2011-0029758 (S. C. P.).

\appendix

%

\section*{Appendix: Cosmological observables with NM coupling}

In this appendix, we derive analytic expressions of cosmological observables following the $\delta$N-formalism \cite{SS}.
We explicitly present the generic results for product-separable potential, $U_1(\phi)U_2(\chi)$, and sum-separable potential, $U_1(\phi)+U_2(\chi)$, with NM coupling. The results nicely reproduce those for the minimal case ~\cite{VW,BW,CHB} when $K=0$ and also those in Refs. \cite{LG} where $e^{2b(\phi)}=1-e^{-2\phi/\sqrt{6}}$ is assumed.

\begin{itemize}

\item{Product-separable potential: $U_1(\phi)U_2(\chi)$}\label{sec:deltaN}

If the potential in Einstein frame (also in Jordan frame) is product-separable, the number of e-foldings is given by
\begin{equation}
N=-\int_{*}^{\rm{e}}\frac{U_{1}}{U_{1}^{\prime}}d\phi=-\int_{*}^{\rm{e}}e^{2b(\phi)}\frac{U_{2}}{U_{2}^{\prime}}d\chi.
\end{equation}

Using the constant of motion along the trajectory \cite{BW},
$C_{\rm{t}}=-\int e^{-2b}\frac{U_{1}}{U_{1}^{\prime}}d\phi+\int \frac{U_{2}}{U_{2}^{\prime}}d\chi$,
we could read out first and second derivatives of $N$ as follows:
\begin{eqnarray}\label{Np}
\frac{\partial N}{\partial \phi_{*}}&=&\frac{s^{\phi}}{\sqrt{2\epsilon_{*}^{\phi}}}\left(1-\frac{\epsilon_{\rm{e}}^{\chi}}{\epsilon_{\rm{e}}}e^{2b_{\rm{e}}-2b_{*}}\right),\nonumber \\
\frac{\partial N}{\partial \chi_{*}}&=&\frac{s^{\chi}}{\sqrt{2\epsilon_{*}^{\chi}}}\left(\frac{\epsilon_{\rm{e}}^{\chi}}{\epsilon_{\rm{e}}}e^{2b_{\rm{e}}-b_{*}}\right),\\
\frac{\partial^{2}N}{\partial \phi_{*}^{2}}&=&
\left(1-\frac{\eta_{*}^{\phi}}{2\epsilon_{*}^{\phi}}\right)\left(1-\frac{\epsilon_{\rm{e}}^{\chi}}{\epsilon_{\rm{e}}}e^{2b_{\rm{e}}-2b_{*}}\right)+\frac{1}{2}s^{b}s^{\phi}\sqrt{\frac{\epsilon_{*}^{b}}{\epsilon_{*}^{\phi}}}\frac{\epsilon_{\rm{e}}^{\chi}}{\epsilon_{\rm{e}}}e^{2b_{\rm{e}}-2b_{*}}+e^{4b_{\rm{e}}-4b_{*}}\frac{1}{\epsilon_{*}^{\phi}}\mathcal{C},\nonumber \\
\frac{\partial^{2}N}{\partial \chi_{*}^{2}}&=&
e^{2b_{\rm{e}}}\left(1-\frac{\eta_{*}^{\chi}}{2\epsilon_{*}^{\chi}}\right)\frac{\epsilon_{\rm{e}}^{\chi}}{\epsilon_{\rm{e}}}+e^{4b_{\rm{e}}-2b_{*}}\frac{1}{\epsilon_{*}^{\chi}}\mathcal{C},\nonumber \\
\frac{\partial^{2}N}{\partial \phi_{*} \partial \chi_{*}}&=&-s^{\phi}s^{\chi}\frac{1}{\sqrt{\epsilon_{*}^{\phi}\epsilon_{*}^{\chi}}}e^{4b_{\rm{e}}-3b_{*}}\mathcal{C},\nonumber
\end{eqnarray}
where $\mathcal{C}\equiv \frac{\epsilon_{\rm{e}}^{\chi}\epsilon_{\rm{e}}^{\phi}}{\epsilon_{\rm{e}}^{2}}\left[\eta_{\rm{e}}^{ss}-4\frac{\epsilon_{\rm{e}}^{\phi}\epsilon_{\rm{e}}^{\chi}}{\epsilon_{\rm{e}}}-\frac{1}{2}s^{b}s^{\phi}\frac{(\epsilon_{\rm{e}}^{\chi})^{2}}{\epsilon_{\rm{e}}}\sqrt{\frac{\epsilon_{\rm{e}}^{b}}{\epsilon_{\rm{e}}^{\phi}}}\,\right]$ and $
\eta^{ss}\equiv\left(\eta^{\phi}\epsilon^{\chi}+\eta^{\chi}\epsilon^{\phi}\right)/\epsilon$ are used\footnote{In $\mathcal{C}$ and $\mathcal{B}_{S}$, the terms $\sqrt{\epsilon^{b}_{\rm{e}}/\epsilon^{\phi}_{\rm{e}}}$ and $\sqrt{\epsilon^{b}_{\rm{e}}\epsilon^{\phi}_{\rm{e}}}$ are read $\sqrt{\epsilon^{b}_{*}/\epsilon^{\phi}_{*}}$ and $\sqrt{\epsilon^{b}_{*}\epsilon^{\phi}_{*}}$, respectively in \cite{CHB}. Also the signs in $\tilde{\alpha}_{u}$ and $\tilde{\alpha}_{v}$ below Eq.~\eqref{eq:COnmsum} are opposite in \cite{CHB}.}.

The power spectrum is given by
\begin{eqnarray}\label{eq:CPS}
\mathcal{P}_{\zeta}=\mathcal{P}_{*}\sum_{I,J}N_{,I}N_{,J}\mathcal{G}^{IJ},
\end{eqnarray}
where $\mathcal{P}_{*}=(H_{*}/2\pi)^{2}$ and $\mathcal{G}^{IJ}$ is defined in such a way that the multi-field action in the Einstein frame is written as of the form of
\begin{eqnarray}
\frac{{\cal L}}{\sqrt{-g}}=-\frac{R}{2}+\frac{1}{2}\sum_{I,J}\mathcal{G}_{IJ}\partial_{\mu}\varphi^{I}\partial^{\mu}\varphi^{J}-U(\varphi^I),
\end{eqnarray}
with $\varphi^I$'s as inflaton fields.
Substituting Eq.~\eqref{Np} into Eq.~\eqref{eq:CPS}, one obtains
\begin{eqnarray}\label{ps}
\mathcal{P}_{\zeta}=\frac{1}{2}\left(\frac{H_{*}}{2\pi}\right)^{2}e^{4b_{\rm{e}}-4b_{*}}\left(\frac{u^{2}\alpha_{p}^{2}}{\epsilon_{*}^{\phi}}+\frac{v^{2}}{\epsilon_{*}^{\chi}}\right),
\end{eqnarray}
where
$u\equiv \epsilon_{\rm{e}}^{\phi}/\epsilon_{\rm{e}}$,  $v\equiv \epsilon_{\rm{e}}^{\chi}/\epsilon_{\rm{e}}$ and $\alpha_{p} \equiv e^{-2b_{\rm{e}}+2b_{*}}\left[1+\frac{\epsilon_{\rm{e}}^{\chi}}{\epsilon_{\rm{e}}^{\phi}}\left(1-e^{2b_{\rm{e}}-2b_{*}}\right)\right]$.

The spectral index, $n_{\zeta}-1\equiv\frac{d\ln\mathcal{P}_{\zeta}}{d\ln k}$ \cite{SS}, is 
\begin{eqnarray}\label{si}
n_{\zeta}-1&=&-2\epsilon_{*}-2\frac{1+N_{,a}\left(\frac{1}{3}R^{abcd}U_{,b}U_{,c}/U^{2}-U^{;ad}/U\right)N_{,d}}{\mathcal{G}^{LM}N_{,L}N_{,M}} \nonumber \\
&=&-2\epsilon_{*}-\frac{4e^{-4b_{\rm{e}}+4b_{*}}}{u^{2}\alpha^{2}/\epsilon_{*}^{\phi}+v^{2}/\epsilon_{*}^{\chi}}-\frac{1}{12}\frac{[\sqrt{\epsilon_{*}^{\chi}/\epsilon_{*}^{\phi}}u\alpha-\sqrt{\epsilon_{*}^{\phi}/\epsilon_{*}^{\chi}}v]^{2}}{u^{2}\alpha^{2}/\epsilon_{*}^{\phi}+v^{2}/\epsilon_{*}^{\chi}}(\eta_{*}^{b}+2\epsilon_{*}^{b})\\
&&+\frac{2}{u^{2}\alpha^{2}/\epsilon_{*}^{\phi}+v^{2}/\epsilon_{*}^{\chi}}
\left[
\frac{u^{2}\alpha^{2}}{\epsilon_{*}^{\phi}}\eta_{*}^{\phi}+\frac{v^{2}}{\epsilon_{*}^{\chi}}\eta_{*}^{\chi}
+4uv\alpha
+\frac{s^{\phi}s^{b}}{2}v\sqrt{\epsilon_{*}^{\phi}\epsilon_{*}^{b}}
\left(\frac{v}{\epsilon_{*}^{\chi}}-\frac{2u\alpha}{\epsilon_{*}^{\phi}}\right)
\right],\nonumber
\end{eqnarray}
where the semicolon denotes the covariant derivative in the scalar field space.

The tensor-to-scalar ratio, $r\equiv\frac{\mathcal{P}_{g}}{\mathcal{P}_{\zeta}}$, with $\mathcal{P}_{g}=8\mathcal{P}_{*}$ is 
\begin{eqnarray}\label{ttsr}
r=\frac{16e^{-4b_{\rm{e}}+4b_{*}}}{u^{2}\alpha_{p}^{2}/\epsilon_{*}^{\phi}+v^{2}/\epsilon_{*}^{\chi}}.
\end{eqnarray}
%
Since the scale-dependent part of the non-linearity parameter is too small to be detectable \cite{VW,CHB}, we just take into account the scale-independent part only:
\begin{eqnarray}\label{fnl}
\frac{6}{5}f_{\rm{NL}}^{(4)}&=&
\frac{2e^{-2b_{\rm{e}}+2b_{*}}}{(u^{2}\alpha_{p}^{2}/\epsilon_{*}^{\phi}+v^{2}/\epsilon_{*}^{\chi})^{2}}
\Bigg[
\frac{u^{3}\alpha_{p}^{3}}{\epsilon_{*}^{\phi}}\left(1-\frac{\eta_{*}^{\phi}}{2\epsilon_{*}^{\phi}}\right)
+\frac{v^{3}}{\epsilon_{*}^{\chi}}\left(1-\frac{\eta_{*}^{\chi}}{2\epsilon_{*}^{\chi}}\right) \nonumber \\
&&+\frac{s^{\phi}s^{b}}{2}\frac{vu^{2}\alpha_{p}^{2}}{(\epsilon_{*}^{\phi})^{2}}\sqrt{\epsilon_{*}^{b}\epsilon_{*}^{\phi}}
+\left(\frac{u\alpha_{p}}{\epsilon_{*}^{\phi}}-\frac{v}{\epsilon_{*}^{\chi}}\right)^{2}e^{2b_{\rm{e}}-2b_{*}}\mathcal{C}
\Bigg].
\end{eqnarray}

When $b=0$, then Eqs.~\eqref{ps}--\eqref{fnl} reduce to the results of minimally coupled case~\cite{BW,CHB}.
In terms of the dimensionless angle $\theta$, the results of non-minimally coupled case in Eqs.~\eqref{ps}--\eqref{fnl} are those in Eqs.~\eqref{eq:COinT}--\eqref{eq:obs}.


\item{Sum-separable potential: $U_1(\phi)+U_2(\chi)$}\label{sec:deltaNsum}

The number of e-foldings for the case of sum potential, $U(\phi,\chi)=U_{1}(\phi)+U_{2}(\chi)$, is given by
\begin{eqnarray}\label{eq:NOE}
N(\phi_{*},\chi_{*})
=-\int_{*}^{e}\frac{U_{1}}{U_{1}^{\prime}}\,d\phi-\int_{*}^{e}\frac{U_{2}}{U_{2}^{\prime}} e^{2b(\phi)} \,d\chi.
\end{eqnarray}

Using the constant of motion along the trajectory, $C_{\rm{t}} = -\int e^{-2b(\phi)}\frac{1}{U_{1}^{\prime}}d\phi +\int \frac{1}{U_{2}^{\prime}}d\chi$, we could read out first and second derivatives of $N$ as follows:
\begin{eqnarray}\label{eq:N1}
\frac{\partial N}{\partial \phi_{*}}&=&\frac{s^{\phi}}{\sqrt{2\epsilon_{*}^{\phi}}} \frac{U_{1*}+Z_{\rm{e}}}{U_{*}}-\mathcal{I},\nonumber \\
\frac{\partial N}{\partial \chi_{*}}&=&\frac{s^{\chi}}{\sqrt{2\epsilon_{*}^{\chi}}} \frac{U_{2*}-Z_{\rm{e}}}{U_{*}}e^{b_{*}}-\mathcal{J}, \nonumber \\
\frac{\partial^{2}N}{\partial \phi_{*}^{2}}&=&1-\frac{\eta_{*}^{\phi}}{2\epsilon_{*}^{\phi}}\frac{U_{1*}+Z_{\rm{e}}}{U_{*}}+ \frac{s^{\phi}}{U_{*}\sqrt{2\epsilon_{*}^{\phi}}}\frac{\partial Z_{\rm{e}}}{\partial \phi_{*}}-\frac{\partial \mathcal{I}}{\partial \phi_{*}}, \nonumber \\
\frac{\partial^{2}N}{\partial \chi_{*}^{2}}&=&e^{2b_{*}}\left(1-\frac{\eta_{*}^{\chi}}{2\epsilon_{*}^{\chi}}\frac{U_{2*}-Z_{\rm{e}}}{U_{*}}- \frac{s^{\chi}e^{-b_{*}}}{U_{*}\sqrt{2\epsilon_{*}^{\chi}}}\frac{\partial Z_{\rm{e}}}{\partial \chi_{*}}\right)-\frac{\partial \mathcal{J}}{\partial \chi_{*}}, \nonumber \\
\frac{\partial^{2}N}{\partial\chi_{*}\partial\phi_{*}}&=&\frac{s^{\phi}}{U_{*}\sqrt{2\epsilon_{*}^{\chi}}}\frac{\partial Z_{\rm{e}}}{\partial \chi_{*}}-\frac{\partial \mathcal{I}}{\partial \chi_{*}}, \nonumber \\
\frac{\partial^{2}N}{\partial\phi_{*}\partial\chi_{*}}&=&s^{\chi}e^{b_{*}}\left( \frac{1}{2}s^{b}\sqrt{\frac{\epsilon_{*}^{b}}{\epsilon_{*}^{\chi}}}\frac{U_{2*}-Z_{\rm{e}}}{U_{*}}-\frac{1}{U_{*}\sqrt{2\epsilon_{*}^{\chi}}}\frac{\partial Z_{\rm{e}}}{\partial \phi_{*}} \right)-\frac{\partial \mathcal{J}}{\partial \phi_{*}}.
\end{eqnarray}
where some useful parameters are introduced as $Z_{\rm{e}}\equiv \frac{U_{2\rm{e}}\epsilon_{\rm{e}}^{\phi}-U_{1\rm{e}}\epsilon_{\rm{e}}^{\chi}}{\epsilon_{\rm{e}}}e^{2b_{\rm{e}}-2b_{*}}$, 
$\mathcal{I} \equiv \int_{*}^{e}\frac{U_{2}}{U_{2}^{\prime}}2b^{\prime}e^{2b}\frac{\partial \phi}{\partial \phi_{*}}d\chi$, and  $\mathcal{J} \equiv \int_{*}^{e}\frac{U_{2}}{U_{2}^{\prime}}2b^{\prime}e^{2b}\frac{\partial \phi}{\partial \chi_{*}}d\chi$.

Cosmological observables, including power spectrum, spectral index, tensor-to-scalar ratio, and the non-linearity parameter, are then given by
\begin{eqnarray}\label{eq:COnmsum}
\mathcal{P}_{\zeta}&=&
\mathcal{P}_{*}\left( \frac{u^{2}}{2\epsilon_{*}^{\phi}}\tilde{\alpha}_{u}^{2}+\frac{v^{2}}{2\epsilon_{*}^{\chi}}\tilde{\alpha}_{v}^{2} \right), \nonumber \\
n_{\zeta}-1&=&-2\epsilon_{*}-\frac{4}{(u^{2}\tilde{\alpha}_{u}^{2}/\epsilon_{*}^{\phi}+v^{2}\tilde{\alpha}_{v}^{2}/\epsilon_{*}^{\chi})}-\frac{1}{12}\left(\eta_{*}^{b}+2\epsilon_{*}^{b}\right)\frac{[u\tilde{\alpha}_{u}\sqrt{\epsilon_{*}^{\chi}/\epsilon_{*}^{\phi}}-v\tilde{\alpha}_{v}\sqrt{\epsilon_{*}^{\phi}/\epsilon_{*}^{\chi}}\,]^{2}}{(u^{2}\tilde{\alpha}_{u}^{2}/\epsilon_{*}^{\phi}+v^{2}\tilde{\alpha}_{v}^{2}/\epsilon_{*}^{\chi})}\nonumber \\
&&+2\frac{(\eta_{*}^{\phi}u^{2}\tilde{\alpha}_{u}^{2}/\epsilon_{*}^{\phi}+\eta_{*}^{\chi}v^{2}\tilde{\alpha}_{v}^{2}/\epsilon_{*}^{\chi})}{(u^{2}\tilde{\alpha}_{u}^{2}/\epsilon_{*}^{\phi}+v^{2}\tilde{\alpha}_{v}^{2}/\epsilon_{*}^{\chi})}-s^{\phi}s^{b}v\tilde{\alpha}_{v}\sqrt{\epsilon_{*}^{b}\epsilon_{*}^{\phi}}\frac{(2u\tilde{\alpha}_{u}/\epsilon_{*}^{\phi}-v\tilde{\alpha}_{v}/\epsilon_{*}^{\chi})}{(u^{2}\tilde{\alpha}_{u}^{2}/\epsilon_{*}^{\phi}+v^{2}\tilde{\alpha}_{v}^{2}/\epsilon_{*}^{\chi})}, \nonumber \\
r&=&\frac{16}{u^{2}\tilde{\alpha}_{u}^{2}/\epsilon_{*}^{\phi}+v^{2}\tilde{\alpha}_{v}^{2}/\epsilon_{*}^{\chi}}, \nonumber \\
f_{\rm{NL}}^{(4)}&=&\frac{5}{6}\frac{2}{\left(u^{2}\tilde{\alpha}_{u}^{2}/\epsilon_{*}^{\phi}+v^{2}\tilde{\alpha}_{v}^{2}/\epsilon_{*}^{\chi}\right)^{2}}\Bigg[
\frac{u^{2}\tilde{\alpha}_{u}^{2}}{\epsilon_{*}^{\phi}}\left( 1-\frac{\eta_{*}^{\phi}}{2\epsilon_{*}^{\phi}}u-\frac{\partial \mathcal{I}}{\partial \phi_{*}}+\frac{\mathcal{C}_{S}}{\epsilon_{*}^{\phi}} \right)+\frac{v^{2}\tilde{\alpha}_{v}^{2}}{\epsilon_{*}^{\chi}}\left( 1-\frac{\eta_{*}^{\chi}}{2\epsilon_{*}^{\chi}}v-e^{-2b_{*}}\frac{\partial \mathcal{J}}{\partial \chi_{*}} \right) \nonumber\\
&&+\left( \frac{u\tilde{\alpha}_{u}}{\epsilon_{*}^{\phi}}-\frac{v\tilde{\alpha}_{v}}{\epsilon_{*}^{\chi}} \right)^{2}\left(\mathcal{A}_{S}+\mathcal{B}_{S}\right)-2s^{\phi}s^{\chi}\frac{u\tilde{\alpha}_{u}}{\sqrt{\epsilon_{*}^{\phi}}}\frac{v\tilde{\alpha}_{v}}{\sqrt{\epsilon_{*}^{\chi}}}e^{-b_{*}}\frac{\partial \mathcal{I}}{\partial \chi_{*}}
\Bigg],
\end{eqnarray}
where $\mathcal{P}_{*}\equiv \left(H_{*}/2\pi\right)^{2}$, $u \equiv (U_{1*}+Z_{e})/U_{*}$, $v \equiv (U_{2*}-Z_{e})/U_{*}$, $\tilde{\alpha}_{u}\equiv 1-s^{\phi}\frac{\sqrt{2\epsilon_{*}^{\phi}}}{u}\mathcal{I}$, and $\tilde{\alpha}_{v}\equiv 1-s^{\chi}\frac{\sqrt{2\epsilon_{*}^{\chi}}}{v}e^{-b_{*}}\mathcal{J}$.

It is useful to use $s^{\phi}\sqrt{2\epsilon_{*}^{\phi}}\frac{\partial Z_{\rm{e}}}{\partial \phi_{*}}=2U_{*}\left( \mathcal{A}_{S}+\mathcal{B}_{S}+\mathcal{C}_{S} \right)$ and $s^{\chi}\sqrt{2\epsilon_{*}^{\chi}}e^{-b_{*}}\frac{\partial Z_{\rm{e}}}{\partial \chi_{*}}=-2U_{*}\left( \mathcal{A}_{S}+\mathcal{B}_{S} \right)$
where
\begin{eqnarray}
&&\mathcal{A}_{S} \equiv -\frac{U_{\rm{e}}^{2}}{U_{*}^{2}}\frac{\epsilon_{\rm{e}}^{\phi}\epsilon_{\rm{e}}^{\chi}}{\epsilon_{\rm{e}}}\left( 1-\frac{\eta_{\rm{e}}^{ss}}{\epsilon_{\rm{e}}}-\frac{1}{2}s^{b}s^{\phi}\frac{\epsilon_{\rm{e}}^{\chi}}{\epsilon_{\rm{e}}^{2}}\sqrt{\epsilon_{\rm{e}}^{b}\epsilon_{\rm{e}}^{\phi}}\, \right)e^{4b_{\rm{e}}-4b_{*}},\\ 
&&\mathcal{B}_{S} \equiv \frac{s^{b}s^{\phi}}{2}\frac{\epsilon_{\rm{e}}^{\chi}}{\epsilon_{\rm{e}}}\frac{U_{\rm{e}}}{U_{*}^{2}}\sqrt{\epsilon_{\rm{e}}^{b}\epsilon_{\rm{e}}^{\phi}}Z_{\rm{e}}e^{2b_{\rm{e}}-2b_{*}},\\ 
&&\mathcal{C}_{S} \equiv -\frac{1}{2}s^{b}s^{\phi}\frac{Z_{\rm{e}}}{U_{*}}\sqrt{\epsilon_{*}^{b}\epsilon_{*}^{\phi}},\\ 
&&\eta^{ss} \equiv \frac{\eta^{\phi}\epsilon^{\chi}+\eta^{\chi}\epsilon^{\phi}}{\epsilon}.
\end{eqnarray}

When $b=0$, then the expressions in Eq.~\eqref{eq:COnmsum} reduce to the results of minimally coupled case in Ref.~\cite{VW}.

\end{itemize}


\begin{thebibliography}{99}

\bibitem{inflation}
  A.~H.~Guth,
  ``The inflationary Universe: A possible solution to the horizon and flatness problems,''
  Phys.\ Rev.\ D {\bf 23}, 347 (1981);
  A.~D.~Linde,
  ``A new inflationary Universe scenario: A possible solution of the horizon, flatness, homogeneity, isotropy and primordial monopole problems,''
  Phys.\ Lett.\ B {\bf 108}, 389 (1982);
  A.~Albrecht and P.~J.~Steinhardt,
  ``Cosmology for grand unified theories with radiatively induced symmetry breaking,''
  Phys.\ Rev.\ Lett.\  {\bf 48}, 1220 (1982).
  
  
\bibitem{WMAP7}
  E.~Komatsu {\it et al.}  [WMAP Collaboration],
  ``Seven-year Wilkinson Microwave Anisotropy Probe (WMAP) observations: Cosmological interpretation,''
  Astrophys.\ J.\ Suppl.\  {\bf 192}, 18 (2011)
  [arXiv:1001.4538 [astro-ph.CO]].
  
  
\bibitem{WMAP9} 
  C.~L.~Bennett, D.~Larson, J.~L.~Weiland, N.~Jarosik, G.~Hinshaw, N.~Odegard, K.~M.~Smith and R.~S.~Hill {\it et al.},
  ``Nine-year Wilkinson Microwave Anisotropy Probe (WMAP) observations: Final maps and results,''
  arXiv:1212.5225 [astro-ph.CO];
  G.~Hinshaw, D.~Larson, E.~Komatsu, D.~N.~Spergel, C.~L.~Bennett, J.~Dunkley, M.~R.~Nolta and M.~Halpern {\it et al.},
  ``Nine-year Wilkinson Microwave Anisotropy Probe (WMAP) observations: Cosmological parameter results,''
  arXiv:1212.5226 [astro-ph.CO].
  
  
  
  
\bibitem{Planck1} 
  P.~A.~R.~Ade {\it et al.}  [Planck Collaboration],
  ``Planck 2013 results. XXIV. Constraints on primordial non-Gaussianity,''
  arXiv:1303.5084 [astro-ph.CO].
  
\bibitem{Planck2} 
  P.~A.~R.~Ade {\it et al.}  [Planck Collaboration],
  ``Planck 2013 results. XVI. Cosmological parameters,''
  arXiv:1303.5076 [astro-ph.CO].
  
\bibitem{Planck3} 
  P.~A.~R.~Ade {\it et al.}  [Planck Collaboration],
  ``Planck 2013 results. XXII. Constraints on inflation,''
  arXiv:1303.5082 [astro-ph.CO].
  
\bibitem{EP} 
  D.~A.~Easson and B.~A.~Powell,
  ``The degeneracy problem in non-canonical inflation,''
  arXiv:1212.4154 [astro-ph.CO].
  
  
  
\bibitem{single} 
  V.~Acquaviva, N.~Bartolo, S.~Matarrese and A.~Riotto,
  ``Second order cosmological perturbations from inflation,''
  Nucl.\ Phys.\ B {\bf 667}, 119 (2003)
 [astro-ph/0209156];
 J.~M.~Maldacena,
  ``Non-Gaussian features of primordial fluctuations in single field inflationary models,''
  JHEP {\bf 0305}, 013 (2003)
  [astro-ph/0210603];
  T.~Tanaka, T.~Suyama and S.~Yokoyama,
  ``Use of delta N formalism: Difficulty of generating large local-type non-Gaussianity during inflation,''
  Prog.\ Theor.\ Phys.\ Suppl.\  {\bf 190}, 135 (2011).
  
  

  
  

\bibitem{BS}
  F.~L.~Bezrukov and M.~Shaposhnikov,
  ``The standard model Higgs boson as the inflaton,''
  Phys.\ Lett.\ B {\bf 659}, 703 (2008)
  [arXiv:0710.3755 [hep-th]].
  
\bibitem{PY}
  S.~C.~Park and S.~Yamaguchi,
  ``Inflation by non-minimal coupling,''
  JCAP {\bf 0808}, 009 (2008)
  [arXiv:0801.1722 [hep-ph]].
  
\bibitem{KP} 
  Y.~Kim and S.~C.~Park,
  ``Hyperbolic inflation,''
  Phys.\ Rev.\ D {\bf 83}, 066009 (2011)
  [arXiv:1010.6021 [hep-ph]].
 
  
  
  
  \bibitem{Qiu} 
  T.~Qiu and K.~-C.~Yang,
  ``Non-Gaussianities of single field inflation with non-minimal coupling,''
  Phys.\ Rev.\ D {\bf 83}, 084022 (2011)
  [arXiv:1012.1697 [hep-th]].
  
  
  
\bibitem{Komatsu:1997} 
  E.~Komatsu and T.~Futamase,
  ``Constraints on the chaotic inflationary scenario with a nonminimally coupled 'inflaton' field from the cosmic microwave background radiation anisotropy,''
  Phys.\ Rev.\ D {\bf 58}, 023004 (1998)
  [astro-ph/9711340].
  
\bibitem{Komatsu:1999} 
  E.~Komatsu and T.~Futamase,
  ``Complete constraints on a nonminimally coupled chaotic inflationary scenario from the cosmic microwave background,''
  Phys.\ Rev.\ D {\bf 59}, 064029 (1999)
  [astro-ph/9901127].
  
  
  
\bibitem{Ferrara:2010yw} 
  S.~Ferrara, R.~Kallosh, A.~Linde, A.~Marrani and A.~Van Proeyen,
  ``Jordan frame supergravity and inflation in NMSSM,''
  Phys.\ Rev.\ D {\bf 82}, 045003 (2010)
  [arXiv:1004.0712 [hep-th]].
  
  
\bibitem{VW}
  F.~Vernizzi and D.~Wands,
  ``Non-gaussianities in two-field inflation,''
  JCAP {\bf 0605}, 019 (2006)
  [astro-ph/0603799];
  
\bibitem{CHB}
  K.~-Y.~Choi, L.~M.~H.~Hall and C.~van de Bruck,
  ``Spectral running and non-Gaussianity from slow-roll inflation in generalised two-field models,''
  JCAP {\bf 0702}, 029 (2007)
  [astro-ph/0701247].
  
  
  
\bibitem{chaotic} 
  A.~D.~Linde,
  ``Chaotic inflation,''
  Phys.\ Lett.\ B {\bf 129}, 177 (1983).

  

\bibitem{natural} 
  K.~Freese, J.~A.~Frieman and A.~V.~Olinto,
  ``Natural inflation with pseudo - Nambu-Goldstone bosons,''
  Phys.\ Rev.\ Lett.\  {\bf 65}, 3233 (1990).


  
  
  
  
  
  
\bibitem{SS}
  M.~Sasaki and E.~D.~Stewart,
  ``A general analytic formula for the spectral index of the density perturbations produced during inflation,''
  Prog.\ Theor.\ Phys.\  {\bf 95}, 71 (1996)
  [astro-ph/9507001].
  
\bibitem{BCH}
  C.~T.~Byrnes, K.~-Y.~Choi and L.~M.~H.~Hall,
  ``Conditions for large non-Gaussianity in two-field slow-roll inflation,''
  JCAP {\bf 0810}, 008 (2008)
  [arXiv:0807.1101 [astro-ph]].
  
\bibitem{BW}
  J.~Garcia-Bellido and D.~Wands,
  ``Metric perturbations in two field inflation,''
  Phys.\ Rev.\ D {\bf 53}, 5437 (1996)
  [astro-ph/9511029].
  
\bibitem{LG}
  J.~-O.~Gong and H.~M.~Lee,
  ``Large non-Gaussianity in non-minimal inflation,''
  JCAP {\bf 1111}, 040 (2011)
  [arXiv:1105.0073 [hep-ph]].
  
\end{thebibliography}
\end{document}